\documentclass[11pt, a4paper]{article}

\usepackage[margin=1truein]{geometry}

\usepackage[T1]{fontenc}
\usepackage{lmodern}
\usepackage{microtype}

\usepackage{amssymb}
\usepackage{bm}
\usepackage{mathtools}
\usepackage{mleftright}\mleftright

\usepackage{amsthm}

\usepackage{algorithm}
\usepackage[noend]{algpseudocode}
\algrenewcommand\algorithmicrequire{\textbf{Input:}}
\algrenewcommand\algorithmicensure{\textbf{Output:}}

\usepackage[
    date=year,
    isbn=false,
    url=false,
    sorting=nyt,
    style=trad-abbrv
]{biblatex}
\usepackage{booktabs}

\usepackage{enumitem}
\setlist[enumerate]{
    label=(\arabic*),
    font=\upshape
}

\usepackage{upref}

\usepackage[
    bookmarksnumbered,
    colorlinks,
    pdfdisplaydoctitle,
    pdfusetitle
]{hyperref}

\usepackage[capitalize,noabbrev]{cleveref}

\crefrangelabelformat{equation}{(#3#1#4\textendash#5#2#6)}

\newcommand{\zcref}[1]{\textup{\cref{#1}}}
\newcommand{\eqrangeref}[1]{\textup{\labelcref{#1}}}

\newtheorem{theorem}{Theorem}
\newtheorem{lemma}[theorem]{Lemma}
\newtheorem{proposition}[theorem]{Proposition}
\newtheorem{corollary}[theorem]{Corollary}
\theoremstyle{remark}
\newtheorem{remark}[theorem]{Remark}

\newcommand{\R}{\mathbb{R}}
\newcommand{\Z}{\mathbb{Z}}
\newcommand{\cS}{\mathcal{S}}
\newcommand{\cF}{\mathcal{F}}

\newcommand{\cI}{\mathcal{I}}

\newcommand{\cK}{\mathcal{K}}
\newcommand{\cB}{\mathcal{B}}
\newcommand{\cC}{\mathcal{C}}
\newcommand{\cM}{\mathcal{M}}

\newcommand{\bM}{\mathbf{M}}
\newcommand{\rmin}{r_{\min}}
\newcommand{\rminp}{r'_{\min}}
\newcommand{\hatrmin}{\hat{r}_{\min}}
\newcommand{\kmax}{k_{\mathrm{max}}}
\newcommand{\Qh}{\hat{Q}}
\newcommand{\symdif}{\mathbin{\triangle}}

\DeclarePairedDelimiter{\prn}{(}{)}
\DeclarePairedDelimiter{\set}{\{}{\}}
\DeclarePairedDelimiterX{\Set}[2]{\{}{\}}{\,#1\mathclose{}\nonscript\;\delimsize|\nonscript\;\mathopen{}#2\,}

\DeclareMathOperator{\supp}{supp}
\DeclareMathOperator*{\argmax}{arg\,max}
\DeclareMathOperator*{\minimize}{minimize}
\DeclareMathOperator{\Ord}{O}

\title{Weighted Matroid Intersection under Minimum Rank Oracle}
\author{
    Taihei Oki\thanks{Institute for Chemical Reaction Design and Discovery (ICReDD), Hokkaido University, Sapporo, Hokkaido, Japan. D3 Center, The University of Osaka, Osaka, Japan. Center for Advanced Intelligence Project, RIKEN, Tokyo, Japan. E-mail: \texttt{oki@icredd.hokudai.ac.jp}.}
}

\begin{document}

\maketitle

\begin{abstract}
    We consider matroid intersection under the minimum rank oracle model, in which two matroids are accessible only through an oracle that returns the minimum of the ranks of a given set in the two matroids.
    By Edmonds' theorem, this oracle determines the matroid intersection polytope, although it does not reveal the exchangeability graphs on which classical algorithms rely.
    Recent work by Bárász, Bérczi, Király, Oki, Yamaguchi, and Yokoi (2025) showed that unweighted matroid intersection is polynomially solvable in this model.

    In this paper, we present a strongly polynomial-time algorithm for weighted matroid intersection in this model.
    The key to our algorithm is a polynomial-size extended formulation of tangent cones of common base polytopes that can be constructed using the minimum rank oracle.
    We also give min-max theorems in terms of the minimum rank function via 1/3-submodularity, and present a strongly polynomial-time algorithm to compute an optimal dual solution.
\end{abstract}

\section{Introduction}\label{sec:intro}

Matroids are among the most fundamental discrete structures in combinatorial optimization.
The independent sets of matroids generalize forests in graphs and linearly independent sets of vectors, and the common independent sets of two matroids on the same ground set generalize matchings in bipartite graphs and branchings in directed graphs (see~\cite{schrijver2003combinatorial}).
Linear optimization over the independent sets and over the common independent sets, possibly subject to cardinality constraints, thus captures a wide range of combinatorial optimization problems, where the latter problem is known as \emph{weighted matroid intersection}.
Both problems have been studied extensively, yielding polyhedral descriptions~\cite{edmonds1970submodular}, min-max theorems~\cite{edmonds1970submodular,frank1981weighted}, and efficient algorithms~\cite{edmonds1971matroids,edmonds1979matroid,lawler1975matroid,frank1981weighted}.

Since a matroid can have exponentially many independent sets, it is typically accessed through an oracle rather than given explicitly.
Standard models are the independence oracle, which tests whether a given set is independent, and the rank oracle, which returns the rank of a given set.
In these polynomially equivalent models, the greedy algorithm solves linear optimization over the independent sets in strongly polynomial time~\cite{edmonds1971matroids}.
Weighted matroid intersection is also solvable in strongly polynomial time when each of the two matroids is given by such an oracle~\cite{edmonds1979matroid,frank1981weighted,lawler1975matroid}; standard algorithms augment a common independent set along shortest paths in the exchangeability graphs.

However, if we view the common independent sets of two matroids as a single discrete structure, the oracles for the individual matroids reveal more than this structure itself.
Motivated by this observation, Bérczi, Király, Yamaguchi, and Yokoi~\cite{berczi2023matroid} initiated the study of matroid intersection under restricted oracles, asking whether the problem remains tractable when independence and rank cannot be queried in each matroid separately.
They introduced the common independence oracle, which tests whether a given set is common independent, and the rank sum oracle, the maximum rank oracle, and the minimum rank oracle, which return the sum, the maximum, and the minimum, respectively, of the ranks of a given set in the two matroids.
The rank sum oracle is motivated by polymatroid matching, and the minimum rank oracle, on which we focus, has a polyhedral motivation explained in the next paragraph.
The difficulty under these restricted oracles is that the oracles do not reveal which of the two matroids forbids a given exchange, so the exchangeability graphs are not directly available.
Nevertheless, they showed that weighted matroid intersection is solvable in strongly polynomial time under the rank sum oracle, and also under the common independence and maximum rank oracles combined.
Under the common independence oracle alone, they solved only special cases, leaving the general case open even for the unweighted problem.

The minimum rank oracle was subsequently studied by Bárász, Bérczi, Király, Oki, Yamaguchi, and Yokoi~\cite{barasz2025matroid}.
By Edmonds' theorem~\cite{edmonds1970submodular}, the minimum rank function gives precisely the inequality description of the matroid intersection polytope, which is the convex hull of the common independent sets.
This oracle is thus natural from a polyhedral viewpoint, raising the question of whether it suffices for efficient linear optimization.
For the unweighted problem, the authors of~\cite{barasz2025matroid} answered this question affirmatively, giving a polynomial-time algorithm that constructs the relevant part of the exchangeability graphs.
For the weighted problem, they solved several special cases, including the case where no circuit of one matroid is contained in a circuit of the other, and gave a fixed-parameter tractable algorithm parameterized by the maximum circuit size of one matroid.
Their approach relies on an exchangeability graph consistent with the oracle answers, which they showed is NP-hard to find in general; the general weighted case was thus left open.

\subsection{Our Contributions}
In this paper, we settle the weighted case under the minimum rank oracle as follows.

\begin{theorem}\label{thm:main}
    There is a strongly polynomial-time algorithm that, given the minimum rank oracle of two matroids on a common ground set $V$ and a weight vector $w\in\R^V$, finds a maximum-weight common independent set of size~$k$ for each $k=0,1,\dots,\kmax$, where $\kmax$ is the maximum size of a common independent set.
    The algorithm makes $\Ord(\kmax^3 n^2)$ oracle calls, where $n \coloneqq |V|$.
\end{theorem}

Unlike the algorithm of~\cite{barasz2025matroid} for the unweighted problem, ours never reconstructs an exchangeability graph.
Instead, we use the minimum rank oracle to compute a polynomial-size extended formulation of the tangent cone of the common base polytope at a common base.
The inequalities defining the tangent cone correspond to the subsets of $V$ that are stable in the exchangeability graph of at least one of the two matroids at this base.
If these two graphs were known, the weighted Kőnig--Egerváry theorem would directly yield an extended formulation.
Although the minimum rank oracle does not reveal them, it does determine the union of their stable-set families.
Our key lemma decomposes this union into polynomially many families, each obtained from the stable sets of a bipartite graph that can be computed with the oracle; applying the weighted Kőnig--Egerváry theorem to each of these bipartite graphs yields the extended formulation.
Since all coefficients of the extended formulation are in $\set{0,\pm1}$, linear programs over it can be solved in strongly polynomial time by Tardos' algorithm~\cite{tardos1986strongly}.
This allows us to compute maximum-weight common independent sets of sizes $k=0,1,\dots,\kmax$ in turn.

We also give min-max theorems for weighted matroid intersection in terms of the minimum rank function, which extend that for the unweighted case in~\cite{barasz2025matroid} and are variants of the classical weight-splitting theorem~\cite{frank1981weighted}.
More generally, we prove these theorems for \emph{1/3-submodular functions}~\cite{berczi2008variations,mizutani2026weakly}, i.e., set functions such that, among any three subsets, some two satisfy the submodular inequality; the minimum of two submodular functions such as the minimum rank function is 1/3-submodular.
Moreover, we compute an optimal dual solution for the minimum rank function case in strongly polynomial time.
Minimizing a general 1/3-submodular function requires exponentially many function value queries~\cite{berczi2008variations}, and so does linear optimization over the corresponding polytope, even for the minimum of a matroid rank function and a 2-polymatroid rank function~\cite{barasz2025matroid}.
In contrast, our results show that minimum rank functions of two matroids is a tractable special case of 1/3-submodular functions.

\subsection{Organization}
\zcref{sec:prelim} gives preliminaries, \zcref{sec:tangent} constructs the extended formulation, and \zcref{sec:algorithm} presents the algorithm and proves \zcref{thm:main}.
\zcref{sec:dual} presents the min-max theorems and an algorithm for the dual problem.

\section{Preliminaries}\label{sec:prelim}

\paragraph{Notation.}
Let $\Z$, $\R$, and $\R_{\ge0}$ be the sets of integers, reals, and nonnegative reals, respectively.
Throughout this paper, let $V$ be a finite ground set of size $n$.
For $S\subseteq V$, we denote $S+v\coloneqq S\cup\set{v}$ when $v\in V\setminus S$ and $S-v\coloneqq S\setminus\set{v}$ when $v\in S$.
For $x\in\R^V$ and $S\subseteq V$, let $x(S)\coloneqq \sum_{v\in S}x_v$, and let $\chi_S\in\set{0,1}^V$ denote the incidence vector of~$S$.

\paragraph{Graphs and hypergraphs.}
For a hypergraph $H$, we write $V(H)$ and $E(H)$ for its vertex and hyperedge sets, respectively.
A vertex subset $S\subseteq V(H)$ is \emph{stable} if it does not include any hyperedge of $H$, and $\cS(H)$ denotes the family of stable sets of $H$.

For a graph $G$, let $\delta_G(v)$ be the set of edges incident with $v \in V(G)$.
For a vertex weight $w\in\R^{V(G)}_{\ge0}$, let $\alpha_w(G) \coloneqq \max\Set{w(S)}{S \in \cS(G)}$ be the maximum $w$-weight of a stable set in $G$.
The following is a weighted variant of the Kőnig--Egerváry theorem.

\begin{proposition}[{see~\cite[Corollary~21.1a]{schrijver2003combinatorial}}]\label{prop:konig}
    For a bipartite graph $G$ and $w\in\R^{V(G)}_{\ge0}$,
    \begin{align}
        w(V(G))-\alpha_w(G)&=\max\Set[\big]{\mu(E(G))}{\mu\in\R^{E(G)}_{\ge0},\ \mu(\delta_G(v))\le w_v\
        (v\in V(G))}.\notag
    \end{align}
\end{proposition}

\paragraph{Matroids.}
A \emph{matroid} $\bM$ consists of a ground set $V$ and a set family $\cI\subseteq2^V$ such that (i) $\emptyset \in \cI$, (ii) $I \in \cI$ and $J \subseteq I$ implies $J \in \cI$, and (iii) for all $I,J\in\cI$ with $|I|<|J|$, there is $v\in J\setminus I$ such that $I+v\in\cI$.
Members of $\cI$ are \emph{independent sets}, and maximal independent sets are \emph{bases}.
The \emph{rank function} $r\colon 2^V \to \Z$ of $\bM$ is defined by $r(S)\coloneqq \max\Set{|I|}{I\subseteq S,\ I\in\cI}$ for $S \subseteq V$.
The \emph{rank} of $\bM$ refers to $r(V)$ and is equal to the cardinality of any base.

Let $\bM_1$ and $\bM_2$ be matroids on a common ground set $V$, and let $r_1$ and $r_2$ be their rank functions.
Assume that both matroids have rank $r$.
Their \emph{matroid intersection polytope} $P$ and \emph{common base polytope} $B$ are the convex hulls of the incidence vectors of their common independent sets and common bases, respectively.
By Edmonds' theorem~\cite{edmonds1970submodular}, they have the following polyhedral description
\begin{align}
    P
    &= \Set[\big]{x\in\R^V_{\ge0}}{x(S)\le r_1(S), \, x(S) \le r_2(S)\ (S\subseteq V)}\notag \\
    &= \Set[\big]{x\in\R^V_{\ge0}}{x(S)\le\rmin(S)\ (S\subseteq V)},\label{eq:mi-polytope}\\
    B
    &= \Set[\big]{x\in P}{x(V)=r}\notag \\
    &= \Set[\big]{x\in\R^V_{\ge0}}{x(S)\le\rmin(S)\ (S\subseteq V),\, x(V)=r},\label{eq:basis-polytope}
\end{align}
where $\rmin(S) \coloneqq \min\set{r_1(S), r_2(S)}$ is called the \emph{minimum rank function}.
Equations~\eqref{eq:mi-polytope} and~\eqref{eq:basis-polytope} show that the minimum rank function determines the polytopes $P$ and $B$.

\paragraph{Computational model and linear programming.}
We count exact arithmetic operations and comparisons over $\R$ at unit cost.
An algorithm runs in \emph{strongly polynomial time} if the number of these operations is bounded by a polynomial in the number of input numbers and, for rational inputs, the binary encoding length of every number arising in the computation is bounded by a polynomial in the binary encoding length of the input.

We use the following algorithmic result on linear programming.

\begin{proposition}[{Tardos~\cite{tardos1986strongly}}]\label{prop:lp}
    Let $A\in\Z^{m\times p}$, $b\in\R^m$, and $c\in\R^p$.
    There is an algorithm that decides whether the linear program $\max\Set{c^\top\xi}{A\xi\le b}$ is infeasible, unbounded, or has an optimal solution, and in the last case computes optimal solutions of it and of its dual $\min\Set{b^\top\pi}{A^\top\pi=c,\ \pi\ge 0}$.
    The number of arithmetic operations and comparisons it performs is bounded by a polynomial in $m$, $p$, and the binary encoding length of $A$.
    Moreover, if $b$ and $c$ are rational, then every number arising in the computation is rational and has binary encoding length bounded by a polynomial in $(A,b,c)$.
\end{proposition}

\zcref{prop:lp} gives a strongly polynomial-time algorithm for linear programming when the entries of $A$ are small integers.
Furthermore, as noted by Megiddo~\cite[p.~240]{megiddo1987complexity}, Tardos' algorithm also applies when $b$ and $c$ belong to an arbitrary ordered field.
In particular, we may use a symbolically perturbed objective $c+\varepsilon c'$, where $\varepsilon$ is a positive infinitesimal.
Such a perturbation can be handled symbolically with polynomial overhead, so the same complexity guarantees apply to lexicographic optimization.

\paragraph{Tangent cones of polytopes.}
The \emph{tangent cone} of a polytope $P\subseteq\R^n$ at $x \in P$ is the set of $d \in \R^n$ such that $x+\varepsilon d \in P$ for all sufficiently small $\varepsilon>0$.
If $P = \Set{x \in \R^n}{Ax \le b}$ for some $A \in \R^{m \times n}$ and $b \in \R^m$, then the tangent cone is given by $\Set{d \in \R^n}{A'd \le 0}$, where $A'$ is the submatrix of $A$ consisting of the rows corresponding to the constraints that are tight at $x$.
The tangent cone is also the conic hull of the vectors $v-x$, where $v$ ranges over the vertices of $P$.

\section{Extended Formulation of Tangent Cones}\label{sec:tangent}

In this section, we construct a polynomial-size extended formulation of tangent cones of the common base polytope that can be computed with the minimum rank oracle.
Throughout this section, fix a common base $B$ of matroids $\bM_1$ and $\bM_2$ over $V$.
Let $r$ be their common rank and $\rmin$ their minimum rank function.

\subsection{Tangent Cones and Exchangeability Graphs}\label{sec:tangent-graphs}

Let $T_B$ be the tangent cone of the common base polytope at $\chi_B$.
For a common independent set $I$, let $\cF_I\coloneqq\Set{S\subseteq V}{\rmin(S)=|I\cap S|}$ be the family of \emph{tight sets} of $I$.
By~\eqref{eq:basis-polytope}, $T_B$ is the set of vectors $d\in\R^V$ satisfying
\begin{align}\label{eq:tangent-cone}
    d(V)=0,\qquad d(S)\le0\quad(S\in\cF_B).
\end{align}
The inequality $d_y\ge0$ $(y\in V\setminus B)$ coming from $x\ge0$ is implied by~\eqref{eq:tangent-cone} with $S=V\setminus\set{y}$.

We rephrase the condition $S\in\cF_B$ by means of the stable sets of \emph{exchangeability graphs}.
For $i\in\set{1,2}$, the \emph{exchangeability graph} $G_B^i$ of $\bM_i$ at $B$ is a bipartite graph with vertex bipartition $\set{B, V\setminus B}$, in which $x\in B$ and $y\in V\setminus B$ are adjacent if and only if $B-x+y$ is a base of $\bM_i$.
Recall that $\cS(\cdot)$ denotes the family of stable sets.

\begin{lemma}\label{lem:rank}
    For $S\subseteq V$, we have $S\in\cF_B$ if and only if $S\symdif B\in\cS(G_B^1)\cup\cS(G_B^2)$.
\end{lemma}

\begin{proof}
    Fix $i\in\set{1,2}$.
    Since $B\cap S$ is independent, $r_i(S)\ge r_i(B \cap S) = |B\cap S|$.
    If $r_i(S)>|B\cap S|$, then $(B\cap S)+y$ is independent for some $y\in S\setminus B$ and extends to a base $B'$ of $\bM_i$.
    Since $y\in B'\setminus B$, the base exchange property gives $x\in B\setminus B'\subseteq B\setminus S$ such that $B-x+y$ is a base, and hence $S\symdif B$ includes the edge $\set{x,y}$ of $G_B^i$.
    Conversely, if $S\symdif B$ includes an edge $\set{x,y}$ of $G_B^i$ with $x\in B$, then $x\notin S$ and $y\in S$, so $(B\cap S)+y\subseteq B-x+y$ is independent and $r_i(S)>|B\cap S|$.
    Hence, $r_i(S)=|B\cap S|$ if and only if $S\symdif B\in\cS(G_B^i)$, implying the claim.
\end{proof}

Define $\sigma_B\colon\R^V\to\R^V$ by $\sigma_B(d)_x \coloneqq -d_x$ for $x \in B$ and $\sigma_B(d)_y\coloneqq d_y$ for $y\in V\setminus B$.
For $d \in \R^V$ with $d(V)=0$ and $h \coloneqq \sigma_B(d)$, we have $h(B)=h(V\setminus B)=d(V \setminus B)$.

\begin{lemma}\label{lem:stable-cone}
    Let $d\in\R^V$ satisfy $d(V)=0$ and $h \coloneqq \sigma_B(d)$.
    Then, $d$ is contained in $T_B$ if and only if $h(S)\le d(V\setminus B)$ for every $S\in\cS(G_B^1)\cup\cS(G_B^2)$.
\end{lemma}

\begin{proof}
    For $S\subseteq V$, we have $d(S)=-h(B\cap S)+h(S\setminus B)=h(S\symdif B)-h(B)=h(S\symdif B)-d(V\setminus B)$.
    Since $S\mapsto S\symdif B$ is a bijection on $2^V$, \zcref{lem:rank} shows that the inequalities in~\eqref{eq:tangent-cone} are equivalent to $h(S')\le d(V\setminus B)$ for $S'\in\cS(G_B^1)\cup\cS(G_B^2)$.
\end{proof}

If $G_B^1$ and $G_B^2$ were known, \zcref{prop:konig} and \zcref{lem:stable-cone}  would already yield a polynomial-size extended formulation of $T_B$: the condition in \zcref{lem:stable-cone} says $\alpha_h(G_B^i)\le d(V\setminus B)$ for $i \in \set{1,2}$.
Thus, by \zcref{prop:konig}, each of these conditions is equivalent to the existence of $\mu^i\in\R^{E_i}_{\ge0}$ with $\mu^i(\delta_i(v))\le h_v$ $(v\in V)$ and $\mu^i(E_i)\ge d(V\setminus B)$, where $E_i \coloneqq E(G_B^i)$ and $\delta_i \coloneqq \delta_{G_B^i}$.
However, neither $G_B^1$ nor $G_B^2$ can be recovered from the minimum rank oracle, and we further rephrase the conditions in \zcref{sec:exch-hypergraphs} below.

\subsection{Exchangeability Hypergraphs}\label{sec:exch-hypergraphs}

Let $H_B$ be the hypergraph defined by $V(H_B) = V$ and
\begin{align}
    E(H_B) = \Set[\big]{X\cup Y}{\begin{array}{l}
        X\subseteq B, Y\subseteq V\setminus B,|X|,|Y| \in \set{1,2}, X \cup Y \notin \cS(G_B^1) \cup \cS(G_B^2)
    \end{array}}.\notag
\end{align}
By \zcref{lem:rank}, $X\cup Y \in E(H_B)$ if and only if $(B\setminus X)\cup Y\notin\cF_B$.
Thus, $H_B$ can be constructed using $\Ord(r^2n^2)$ calls to the minimum rank oracle.
Together with \zcref{lem:stable-cone}, the following lemma gives a description of $T_B$ in terms of $\cS(H_B)$.

\begin{lemma}\label{lem:hypergraph}
    $\cS(H_B)=\cS(G_B^1)\cup\cS(G_B^2)$.
\end{lemma}

\begin{proof}
    If $S\subseteq V$ includes edges $e_1 \in E(G_B^1)$ and $e_2 \in E(G_B^2)$, then $S$ includes the hyperedge $e_1\cup e_2 \in E(H_B)$.
    Conversely, every hyperedge of $H_B$ includes an edge of $G_B^i$ for $i \in \set{1,2}$ by definition.
\end{proof}

Although \zcref{lem:hypergraph} allows us to replace $\cS(G_B^1)\cup\cS(G_B^2)$ in \zcref{lem:stable-cone} with $\cS(H_B)$, this does not immediately yield a polynomial-size linear programming formulation.
We therefore express $\cS(H_B)$ as a union of polynomially many families derived from the stable sets of bipartite graphs that can be constructed using the minimum rank oracle.
Applying \zcref{prop:konig} to these graphs then yields the desired formulation.

For each pair $e=\set{x,y}$ with $x \in B$ and $y \in V \setminus B$, let $\cM_e$ consist of the inclusion-minimal members of $\Set{D\setminus e}{D\in E(H_B)}$.
Let $G_0$ be the bipartite graph with vertex bipartition $\set{B, V\setminus B}$ whose edges are precisely the pairs $e$ for which every member of $\cM_e$ is empty, a singleton, or a pair with one element in $B$ and the other in $V \setminus B$.
The following lemma shows that $G_0$ contains both $G_B^1$ and $G_B^2$ as subgraphs.

\begin{lemma}\label{lem:supergraph}
    $E(G_B^1)\cup E(G_B^2)\subseteq E(G_0)$.
\end{lemma}

\begin{proof}
    Let $e\in E(G_B^1)$; the case of $G_B^2$ is symmetric.
    Let $D\in E(H_B)$ be such that $D\setminus e \in \cM_e$.
    By the definition of $H_B$, $D$ contains an edge $f$ of $G_B^2$.
    Since $e\cup f$ contains an edge of each of $G_B^1$ and $G_B^2$, it is a hyperedge of $H_B$.
    As $f\setminus e=(e\cup f)\setminus e$ is contained in $D\setminus e$, the minimality of $\cM_e$ gives $D\setminus e=f\setminus e$.
    Hence, $D \setminus e$ is empty (if $f=e$), a singleton (if $|f\cap e|=1$), or the edge $f$ of $G_B^2$, which has one endpoint in each part (if $f\cap e=\emptyset$).
    Therefore, $e\in E(G_0)$.
\end{proof}

By \zcref{lem:supergraph,lem:hypergraph}, we have $\cS(G_0)\subseteq\cS(G_B^1)\cap\cS(G_B^2)\subseteq\cS(H_B)$.
To describe the stable sets of $H_B$ that are not stable in $G_0$, we introduce the following bipartite graphs.
For each $e\in F \coloneqq E(G_0)\setminus E(H_B)$, let $X_e\coloneqq \Set{v\in V}{\set{v}\in\cM_e}$ and define a graph $G_e$ by $V(G_e)=V\setminus(X_e\cup e)$ and $E(G_e)=\Set{f\in\cM_e}{|f|=2}$.
Every edge $f\in E(G_e)$ is disjoint from $e$ by definition and from $X_e$ by inclusion-minimality, so both ends of $f$ lie in $V(G_e)$.
Moreover, since $e\in E(G_0)$, each such edge has one endpoint in $B$ and the other in $V\setminus B$.
Thus, $G_e$ is a well-defined bipartite graph.

\begin{lemma}\label{lem:decomposition}
    We have
    \begin{align}\label{eq:decomposition}
        \cS(H_B)&=\cS(G_0)\ \cup\bigcup_{e\in F}\Set{S\cup e}{S\in\cS(G_e)}.
    \end{align}
\end{lemma}
\begin{proof}
    We first claim that, for every $e\in F$ and $S\subseteq V\setminus e$, the set $S\cup e$ is stable in $H_B$ exactly when $S$ is stable in $G_e$.
    Since $e\in E(G_0)$, each member of $\cM_e$ is empty, a singleton $\set{v}$ with $v\in X_e$, or an edge of $G_e$.
    By $e \notin E(H_B)$, we have $\emptyset \notin \cM_e$.
    Since $D\subseteq S\cup e$ is equivalent to $D\setminus e\subseteq S$ for every $D \in E(H_B)$, the set $S\cup e$ is stable in $H_B$ exactly when $S$ contains no member of $\cM_e$, that is, when $S$ avoids $X_e$ and contains no edge of $G_e$.
    This proves the claim.

    By this claim and $\cS(G_0)\subseteq\cS(H_B)$, the right-hand side of~\eqref{eq:decomposition} is included in $\cS(H_B)$.
    Conversely, let $S\in\cS(H_B)$.
    If $S$ contains no edge of $G_0$, then $S\in\cS(G_0)$.
    Otherwise, $S$ contains some $e\in E(G_0)$, which is not a hyperedge of $H_B$ because $S$ is stable in $H_B$.
    Thus, $e\in F$, and $S\setminus e\in\cS(G_e)$ by the claim.
\end{proof}

By \zcref{lem:decomposition}, $\cS(H_B)$ is the union of $\Ord(rn)$ families, each of which is obtained from the stable sets of a bipartite graph.
These graphs can be constructed with $\Ord(r^2n^2)$ calls to the minimum rank oracle and polynomially many further operations: after building $H_B$, we obtain each $\cM_e$ by taking the inclusion-minimal members of $\Set{D\setminus e}{D\in E(H_B)}$.

\subsection{The Extended Formulation}\label{sec:ef}

\zcref{lem:decomposition} yields the following polynomial-size extended formulation of $T_B$.

\begin{theorem}\label{thm:ef}
    Let $\xi=(d,\mu,\nu)$, where $d\in\R^V$, $\mu\in\R^{E(G_0)}$, and $\nu=(\nu^e)_{e\in F}$ with $\nu^e\in\R^{E(G_e)}$ for each $e\in F$.
    The tangent cone $T_B$ is the projection onto the $d$-coordinates of the set of such vectors satisfying $d(V)=0$ and
    \begin{alignat}{2}
        \mu(\delta_{G_0}(v)) &\le h_v &\quad & (v\in V),
        \label{eq:ef-k-capacity}\\
        \mu(E(G_0)) &\ge d(V\setminus B),
        \label{eq:ef-k}\\
        \mu_f &\ge 0 &\quad & (f\in E(G_0)),
        \label{eq:ef-k-nonneg}\\
        \nu^e(\delta_{G_e}(v)) &\le h_v &\quad & (e \in F, v\in V(G_e)),
        \label{eq:ef-q-capacity}\\
        \nu^e(E(G_e)) &\ge d(V\setminus B)-h(X_e)&\quad & (e \in F),
        \label{eq:ef-q}\\
        \nu^e_f &\ge 0 &\quad & (e\in F, f\in E(G_e)),
        \label{eq:ef-q-nonneg}
    \end{alignat}
    where $h \coloneqq \sigma_B(d)$.
    The system consists of $\Ord(r^2n^2)$ homogeneous linear constraints in $\Ord(r^2n^2)$ variables, and all its coefficients are in $\set{0,\pm1}$.
\end{theorem}

\begin{proof}
    Let $d$ satisfy $d(V)=0$.
    We may assume $h\ge0$, as it is implied both by $d\in T_B$ (via~\eqref{eq:tangent-cone} with $S=\set{x}$ and $S=V\setminus\set{y}$) and by~\eqref{eq:ef-k-capacity} and~\eqref{eq:ef-k-nonneg}.
    By \zcref{lem:stable-cone,lem:hypergraph,lem:decomposition}, $d\in T_B$ if and only if (i) $\alpha_h(G_0)\le d(V\setminus B)$ and (ii) $\alpha_h(G_e) + h(e) \le d(V\setminus B)$ for all $e\in F$, where $h$ is restricted to $V(G_e)$ in $\alpha_h(G_e)$.
    Since $h(V)=2d(V\setminus B)$, \zcref{prop:konig} shows that the condition (i) is equivalent to the existence of $\mu$ satisfying~\eqrangeref{eq:ef-k-capacity,eq:ef-k,eq:ef-k-nonneg}.
    Since $e$, $X_e$, and $V(G_e)$ partition $V$, we have $h(V(G_e))-(d(V\setminus B)-h(e))=d(V\setminus B)-h(X_e)$.
    Thus, by \zcref{prop:konig}, the condition (ii) is equivalent to the existence of $\nu^e$ satisfying~\eqrangeref{eq:ef-q-capacity,eq:ef-q,eq:ef-q-nonneg}.

    We have $|F|\le r(n-r)$, and each $G_e$ has at most $r(n-r)$ edges and at most $n$ vertices.
    Hence, the system has $n+|E(G_0)|+\sum_{e\in F}|E(G_e)|=\Ord(r^2n^2)$ variables and $\Ord(n+r(n-r)n+r^2(n-r)^2)=\Ord(r^2n^2)$ constraints.
    For the coefficients, note that $h_v=\pm d_v$ and that $d(V\setminus B)-h(X_e)=h(B\setminus X_e)-h(X_e\setminus B)=-d(B\setminus X_e)-d(X_e\setminus B)$, where the two index sets are disjoint.
\end{proof}

\section{The Algorithm}\label{sec:algorithm}

In this section, we present our matroid intersection algorithm and prove \zcref{thm:main}.

For $i \in \set{1,2}$, let $\bM_i$ be a matroid over $V$ with independent set family $\cI_i$.
Let $\rmin$ be the minimum rank function of $(\bM_1, \bM_2)$, and for $k\ge0$, let $\cI^{(k)}$ be their common independent sets of size $k$.
Let $\kmax$ be the maximum size of a common independent set, which is unknown in advance.
Let $w \in \R^V$ be a weight vector.

\subsection{Lexicographic Tie-Breaking}

Before running the algorithm, we perturb the weight vector $w$ so that the maximum is uniquely attained.
For this, fix an ordering $V=\set{v_1,\ldots,v_n}$ and define $\eta\in\Z^V$ by $\eta_{v_i}\coloneqq 2^i$ for $i \in \set{1, \dotsc, n}$.
Write $\tilde{w}\coloneqq w+\varepsilon\eta$ for the symbolic perturbation of $w$, where $\varepsilon$ is a positive infinitesimal.
Maximizing $\tilde{w}^\top d$ means lexicographically maximizing $(w^\top d,\eta^\top d)$.
Since $\eta(S)\ne\eta(T)$ for distinct $S,T\subseteq V$, the maximizer of $\tilde{w}$ over any nonempty family of subsets of $V$ is unique.

\subsection{Algorithm Description}\label{sec:algorithm-description}

Starting from $I_0^*=\emptyset$, our algorithm successively computes the unique maximizer $I_k^*$ of $\tilde{w}$ over $\cI^{(k)}$ for $k=1,\ldots,\kmax$.
\zcref{alg:stage} describes a stage of the algorithm, which computes $I_{k+1}^*$ from $I_k^*$ or decides that $k=\kmax$.

\begin{algorithm}[t]
    \caption{Computing $I_{k+1}^*$ from $I_k^*$}\label{alg:stage}
    \begin{algorithmic}[1]
        \Require the minimum rank oracle of $(\bM_1, \bM_2)$, $w\in\R^V$, $I_k^* \in \argmax\Set{\tilde{w}(I)}{I \in \cI^{(k)}}$
        \Ensure $I_{k+1}^* \in \argmax\Set{\tilde{w}(I)}{I \in \cI^{(k+1)}}$, or the answer that $k=\kmax$
        \State Let $B \coloneqq I_k^* + \star$ and build $H_B, G_0$, and $G_e$ $(e \in F)$ for $(\bM_1', \bM_2')$
        \State Write $d(V')=0$, \eqrangeref{eq:ef-k-capacity,eq:ef-k,eq:ef-k-nonneg,eq:ef-q-capacity,eq:ef-q,eq:ef-q-nonneg}, and~\eqref{eq:box} as a linear inequality system $A\xi\le b$
        \If{$A\xi\le b$ has no solution}
            \State \Return ``$k=\kmax$''
        \EndIf
        \State Compute an optimal solution $\xi^*=(d^*, \mu^*, \nu^*)$ of
            $\max\Set{\tilde{w}^\top d}{A\xi\le b}$
        \State \Return $I_{k+1}^*\coloneqq \Set{v\in V}{(\chi_B+d^*)_v=1}$
    \end{algorithmic}
\end{algorithm}

For $i \in \set{1,2}$, we construct a matroid $\bM_i'$ from $\bM_i$ as follows: the ground set is $V' \coloneqq V + \star$, where $\star \notin V$ is a new element, and the independent set family is given by
\begin{align}\label{eq:auxiliary}
    \cI_i'\coloneqq \Set{S\subseteq V'}{|S|\le k+1,\ S-\star\in\cI_i}.
\end{align}
That is, $\bM_i'$ is the truncation to rank $k+1$ of the direct sum of $\bM_i$ and the free matroid on $\set{\star}$.
As $|I_k^*|=k$ and $I_k^*\in\cI_i$, the matroid $\bM_i'$ has rank $k+1$, and $I_k^*+\star$ is a common base of $\bM_1'$ and $\bM_2'$.
The minimum rank function of the pair $(\bM_1',\bM_2')$ satisfies
\begin{align}\label{eq:auxiliary-oracle}
    \rminp(S)&=\min\set{k+1,\ \rmin(S \setminus \set{\star})+|S\cap\set{\star}|}\qquad(S\subseteq V'),
\end{align}
where $\rmin$ is the minimum rank function of $(\bM_1, \bM_2)$.
Thus, we can compute $\rminp(S)$ in one call to $\rmin$ for each $S \subseteq V'$.
We extend $w$, $\eta$, and $\tilde{w}$ to $V'$ by setting $w_{\star}=\eta_{\star}=\tilde{w}_{\star}\coloneqq0$.

We apply the arguments in \zcref{sec:tangent} to $\bM_1'$, $\bM_2'$, and the common base $B \coloneqq I_k^* + \star$.
We build the hypergraph $H_B$ and the bipartite graphs $G_0$ and $G_e$ $(e \in F)$ using $\rminp$, and construct the linear system consisting of $d(V')=0$ and~\eqrangeref{eq:ef-k-capacity,eq:ef-k,eq:ef-k-nonneg,eq:ef-q-capacity,eq:ef-q,eq:ef-q-nonneg} for the tangent cone $T_B$ of the common base polytope of $(\bM_1', \bM_2')$ at $B$.
We add to the system the following constraints
\begin{align}\label{eq:box}
    d_{\star}&=-1,\qquad -1\le d_x\le0\ \ (x\in B),\qquad
    0\le d_y\le1\ \ (y\in V'\setminus B).
\end{align}
Let $A\xi\le b$ be the resulting system in variable $\xi=(d,\mu,\nu)$; all entries of $A$ are in $\set{0,\pm1}$.

We maximize $\tilde w^\top d$ subject to $A\xi\le b$ using \zcref{prop:lp}, with $\varepsilon$ handled symbolically.
Since the objective depends only on $d$, which is bounded by~\eqref{eq:box}, an optimal solution exists whenever the system is feasible.
We either determine that the system is infeasible or compute an optimal solution $\xi^*=(d^*, \mu^*, \nu^*)$.
In the former case, we report $k=\kmax$; in the latter case, we return $\Set{v\in V}{(\chi_B+d^*)_v=1}$ as $I_{k+1}^*$.

\subsection{Correctness}
We now prove the correctness of \zcref{alg:stage}.
Let $\cB$ be the family of common bases of $\bM_1'$ and $\bM_2'$.
By~\eqref{eq:auxiliary}, we have $\cB=\Set{J\subseteq V'}{|J|=k+1,\ J-\star\in\cI_1\cap\cI_2}$,
which is partitioned into $\cB_\star\coloneqq \Set{J\in\cB}{\star\in J}$ and $\cI^{(k+1)} = \Set{J\in\cB}{\star\notin J}$.
There is a one-to-one correspondence between $\cB_\star$ and $\cI^{(k)}$ via $\cB_\star \ni J\mapsto J-\star \in \cI^{(k)}$.
As $w_{\star}=\eta_{\star}=0$, we have $w(J)=w(J-\star)$ and $\eta(J)=\eta(J-\star)$ for $J\in\cB_\star$.

Let $Q\coloneqq \Set{d\in T_B}{d_{\star}=-1,\ 0\le\chi_B+d\le1}$ and $\Qh \coloneqq \Set{\xi}{A\xi \le b}$.
By \zcref{thm:ef} and~\eqref{eq:box}, $Q$ is the projection of $\Qh$ onto the $d$-coordinates.

\begin{lemma}\label{lem:feasibility}
    $Q\ne\emptyset$ if and only if $\cI^{(k+1)}\ne\emptyset$.
\end{lemma}

\begin{proof}
    Let $d\in Q$.
    As $d\in T_B$, there are coefficients $\lambda_J\ge0$ $(J\in\cB)$ such that
    \begin{align}\label{eq:conic-representation}
        d&=\sum_{J\in\cB}\lambda_J(\chi_J-\chi_B).
    \end{align}
    Then, ${(\chi_J-\chi_B)}_\star$ is $0$ for $J\in\cB_\star$ and $-1$ for $J\in\cI^{(k+1)}$.
    Hence, $d_{\star}=-1$ implies
    \begin{align}\label{eq:mass}
        \sum_{J\in\cI^{(k+1)}}\lambda_J&=1,
    \end{align}
    and in particular $\cI^{(k+1)}\ne\emptyset$.
    Conversely, for every $J\in\cI^{(k+1)}$, the vector $\chi_J-\chi_B$ lies in $T_B$, its $\star$-coordinate is $-1$, and $\chi_B+(\chi_J-\chi_B)=\chi_J\in{[0,1]}^{V'}$; thus it lies in~$Q$.
\end{proof}

\begin{lemma}\label{lem:augmentation}
    If $\cI^{(k+1)}\ne\emptyset$, then $\chi_{I_{k+1}^*}-\chi_B$ is the unique maximizer of $\tilde{w}^\top d$ over $Q$.
\end{lemma}

\begin{proof}
    The vector $\chi_{I_{k+1}^*}-\chi_B$ belongs to $Q$.
    Let $d\in Q$ and choose $\lambda_J\ge0$ $(J \in \cB)$ satisfying~\eqref{eq:conic-representation} and~\eqref{eq:mass}.
    For $J\in\cB_\star$, we have $J-\star\in\cI^{(k)}$ and $\tilde{w}(J)=\tilde{w}(J-\star)\le\tilde{w}(I_k^*)=\tilde{w}(B)$, with equality only if $J=B$ as $I_k^*$ uniquely maximizes $\tilde{w}$ over $\cI^{(k)}$.
    For $J\in\cI^{(k+1)}$, the maximality of $I_{k+1}^*$ gives $\tilde{w}(J)\le\tilde{w}(I_{k+1}^*)$, with equality only if $J=I_{k+1}^*$.
    Since $\cB$ is the disjoint union of $\cB_\star$ and $\cI^{(k+1)}$, and all $\lambda_J$ are nonnegative, we obtain
    \begin{align}
        \tilde{w}^\top d
        &=\sum_{J\in\cB}\lambda_J(\tilde{w}(J)-\tilde{w}(B))\notag \\
        &\le\sum_{J\in\cI^{(k+1)}}\lambda_J(\tilde{w}(J)-\tilde{w}(B))\notag\\
        &\le (\tilde{w}(I_{k+1}^*)-\tilde{w}(B))
        \sum_{J\in\cI^{(k+1)}}\lambda_J\notag \\
        &=\tilde{w}^\top(\chi_{I_{k+1}^*}-\chi_B).\notag
    \end{align}
    Equality requires $\lambda_J=0$ for all $J\notin\set{B,I_{k+1}^*}$.
    In this case, \eqref{eq:mass} gives $\lambda_{I_{k+1}^*}=1$, and hence $d=\chi_{I_{k+1}^*}-\chi_B$.
\end{proof}

\begin{corollary}\label{cor:stage}
    \zcref{alg:stage} returns ``$k=\kmax$'' if $\cI^{(k+1)} = \emptyset$, and $I_{k+1}^*$ otherwise.
\end{corollary}

\begin{proof}
    Since $Q$ is the projection of $\Qh$ onto the $d$-coordinates, \zcref{lem:feasibility} implies that $\Qh=\emptyset$ if and only if $\cI^{(k+1)}=\emptyset$.
    Assume $\cI^{(k+1)}\ne\emptyset$.
    The symbolic objective $\tilde{w}^\top d$ depends only on $d$, and the projection of $\Qh$ onto the $d$-coordinates is $Q$.
    By \zcref{lem:augmentation}, the $d$-component $d^*$ of the optimal solution computed by \zcref{alg:stage} equals $\chi_{I_{k+1}^*}-\chi_B$.
    Thus, $\chi_B+d^*=\chi_{I_{k+1}^*}$, so the algorithm returns $I_{k+1}^*$.
\end{proof}

We are now ready to prove \zcref{thm:main}.

\begin{proof}[Proof of \zcref{thm:main}]
    By \zcref{cor:stage} and induction on $k$, running \zcref{alg:stage} for $k=0,1,\ldots$ until it reports $k=\kmax$ (or until $k=n$) yields $I_0^*,\ldots,I_{\kmax}^*$, and each $I_k^*$ is a maximum-weight common independent set of size $k$.

    We analyze the number of calls to the minimum rank oracle.
    In stage $k$ of \zcref{alg:stage}, we have $|B|=k+1$ and $|V'\setminus B|=n-k$.
    Building $H_B$ takes $\Ord(k^2 n^2)$ evaluations of $\rminp$, each of which is one oracle call by~\eqref{eq:auxiliary-oracle}.
    No other step queries the oracle.
    Hence, the total number of oracle calls is $\sum_{k=0}^{\kmax}\Ord(k^2 n^2)=\Ord(\kmax^3 n^2)$.

    Next, we check that \zcref{alg:stage} performs polynomially many arithmetic operations.
    In stage $k$, $H_B$ has $\Ord(k^2 n^2)$ hyperedges of size at most four, so the bipartite graphs in \zcref{lem:decomposition} are computed with a polynomial number of operations.
    By \zcref{thm:ef}, the system $A\xi\le b$ has $\Ord(k^2 n^2)$ variables and constraints, and all entries of $A$ and $b$ are in $\set{0,\pm1}$.
    By \zcref{prop:lp}, the feasibility test and the optimization with the symbolic objective take a number of operations polynomial in $n$.
    As there are at most $\kmax+1 = \Ord(n)$ stages, the total number of operations is polynomial in $n$.

    We finally check the bit complexity for rational $w$.
    Tardos' algorithm keeps the encoding lengths of all intermediate numbers polynomial in the input length.
    All other computations are combinatorial or involve only $\eta$ and the constraint data, which have encoding length polynomial in $n$.
    Therefore, the algorithm is strongly polynomial.
\end{proof}

\section{Min--Max Theorems}\label{sec:dual}

In this section, we present a variant of the weight-splitting theorem for weighted matroid intersection involving only the minimum rank function via 1/3-submodularity.
We also show how to compute an optimal weight splitting.

\subsection{1/3-Submodular Functions and Min--Max Theorems}\label{sec:minmax}

Recall that a set function $f \colon 2^V \to \R$ is \emph{submodular} if $f(S)+f(T)\ge f(S \cap T)+f(S \cup T)$ holds for $S,T \subseteq V$.
A set function $f$ is \emph{1/3-submodular}~\cite{berczi2008variations,mizutani2026weakly} if, among any three subsets $S_1, S_2, S_3 \subseteq V$, there is a pair $S_i, S_j$ with $i \ne j$ such that $f(S_i)+f(S_j)\ge f(S_i \cap S_j)+f(S_i \cup S_j)$.
Following the standard terminology on polymatroids, we refer to $f$ as a (\emph{1/3-})\emph{polymatroid rank function} if $f(\emptyset)=0$, $f$ is monotone (i.e.\ $f(S)\le f(T)$ for $S \subseteq T \subseteq V$), and $f$ is (1/3-)submodular.
The minimum of two submodular functions is $1/3$-submodular, as two of any three sets attain the minimum in the same function.
Similarly, the minimum of two polymatroid rank functions, such as the minimum rank function $\rmin$ of two matroids $\bM_1$ and $\bM_2$, is a $1/3$-polymatroid rank function.

For a 1/3-polymatroid rank function $f$, define
\begin{align}
    P(f)&\coloneqq\Set{x\in\R^V_{\ge0}}{x(S)\le f(S)\ (S\subseteq V)},\notag \\
    B(f)&\coloneqq\Set{x\in P(f)}{x(V)=f(V)},\notag
\end{align}
which we call the \emph{1/3-polymatroid} defined by $f$ and its \emph{base polyhedron}.
While $P(f) \ne \emptyset$ as $0 \in P(f)$, the base polyhedron $B(f)$ can be empty\footnote{
    Consider $|V|=2$, $f(S) \coloneqq 0$ for $|S| \le 1$, and $f(V) \coloneqq 1$.
    Then, $P(f) = \set{(0,0)}$ and $B(f) = \emptyset$.
}, unlike the polymatroid case.
These are integral polytopes if $f$ is integer-valued (see~\cite{berczi2008variations}).
Note that $P(\rmin)$ and $B(\rmin)$ are the matroid intersection and common base polytopes of $(\bM_1, \bM_2)$, respectively.

For a set function $f\colon 2^V \to \R$ with $f(\emptyset)=0$, its \emph{Lovász extension} $\hat f\colon\R^V\to\R$ is given by
\begin{align}\label{eq:lovasz}
    \hat f(p)\coloneqq\sum_{k=1}^n p_{v_k}(f(S_k)-f(S_{k-1}))=\sum_{k=1}^{n-1}(p_{v_k}-p_{v_{k+1}})f(S_k)+p_{v_n}f(V),
\end{align}
where $p_{v_1}\ge\cdots\ge p_{v_n}$, $S_k\coloneqq\set{v_1,\ldots,v_k}$, and $S_0\coloneqq\emptyset$.
The function $\hat f$ satisfies $\hat f(p+a\chi_V)=\hat f(p)+af(V)$ for $a\in\R$, $\hat f(\chi_S) = f(S)$ for $S \subseteq V$, and is nondecreasing if $f$ is monotone.
It is well-known that $\hat f$ is convex if and only if $f$ is submodular~\cite{lovasz1983submodular}.

By uncrossing arguments, a min-max theorem for linear optimization over 1/3-polymatroids is given as follows.
For $p\in\R^V$, let $p^+\in\R^V_{\ge0}$ be defined by $p_v^+\coloneqq\max\set{p_v,0}$ for $v\in V$.
We call a pair $(w_1,w_2)$ with $w_1+w_2=w$ a \emph{weight splitting} of $w$.

\begin{theorem}\label{thm:one-third-minmax}
    Let $f\colon2^V\to\R$ be a 1/3-polymatroid rank function and $w\in\R^V$.
    Then,
    \begin{align}\label{eq:function-splitting}
        \max_{x\in P(f)}w^\top x
        &=\min\Set[\big]{\hat f(w_1^+)+\hat f(w_2^+)}{w_1,w_2\in\R^V,\ w_1+w_2=w},
    \end{align}
    and, if $B(f)\ne\emptyset$,
    \begin{align}\label{eq:function-base-splitting}
        \max_{x\in B(f)}w^\top x
        =\min\Set[\big]{\hat f(w_1)+\hat f(w_2)}{w_1,w_2\in\R^V,\ w_1+w_2=w}.
    \end{align}
    If $w$ is integer-valued, both minima are attained by integer-valued vectors.
\end{theorem}

We prove \zcref{thm:one-third-minmax} using the following lemma on the dual of $\max\Set{w^\top x}{x \in P(f)}$ for a 1/3-submodular function $f$.
For $y\in\R^{2^V}$, let $\supp(y)\coloneqq\Set{S\subseteq V}{y_S\ne0}$ denote its support, and define the potential function $\varphi(y)\coloneqq\sum_{S\subseteq V}(n^2-|S|^2)y_S$.
The lemma is essentially due to~\cite{berczi2008variations}; we include a proof in our notation for completeness.

\begin{lemma}\label{lem:two-chain}
    Let $f\colon2^V\to\R$ be $1/3$-submodular and $w\in\R^V$.
    If $y \in \R_{\ge0}^{2^V}$ minimizes $\sum_{S \subseteq V}f(S)y_S$ and then $\varphi(y)$ subject to $\sum_{S \subseteq V}y_S\chi_S\ge w$, then $\supp(y)$ is the union of two chains.
\end{lemma}

\begin{proof}
    Suppose that $\supp(y)$ contains three pairwise incomparable sets $S_1$, $S_2$, and $S_3$.
    By $1/3$-submodularity of $f$, there exist $S_i$ and $S_j$ such that $i \ne j$ and $f(S_i)+f(S_j)\ge f(S_i\cap S_j)+f(S_i\cup S_j)$.
    Decrease $y_{S_i}$ and $y_{S_j}$ by $\delta\coloneqq\min\set{y_{S_i},y_{S_j}}>0$ and increase $y_{S_i\cap S_j}$ and $y_{S_i\cup S_j}$ by $\delta$.
    Since $\chi_{S_i}+\chi_{S_j}=\chi_{S_i\cap S_j}+\chi_{S_i\cup S_j}$, the resulting vector has the same $\sum_{S \subseteq V} y_S\chi_S$ and remains feasible; the primal objective $\sum_{S \subseteq V} f(S)y_S$ does not increase, so it is still a minimizer; and the secondary objective $\varphi(y)$ decreases by $\delta(|S_i\cap S_j|^2+|S_i\cup S_j|^2-|S_i|^2-|S_j|^2)=2\delta|S_i\setminus S_j|\,|S_j\setminus S_i|>0$, a contradiction.
    Hence, $\supp(y)$ has no three pairwise incomparable members, and is covered by two chains by Dilworth's theorem (see~\cite[Section~14.2]{schrijver2003combinatorial}).
\end{proof}

We are now ready to prove \zcref{thm:one-third-minmax}.

\begin{proof}[Proof of \zcref{thm:one-third-minmax}]
    We first show that
    \begin{align}\label{eq:function-covering}
        \max_{x\in P(f)}w^\top x
        =\min\Set[\big]{\hat f(w_1)+\hat f(w_2)}{w_1,w_2\in\R^V_{\ge0},\ w_1+w_2\ge w}.
    \end{align}
    Consider the linear program
    \begin{align}\label{eq:function-dual}
        \minimize_{y \in \R_{\ge 0}^{2^V}} \quad \sum_{S\subseteq V}f(S)y_S \quad
        \mathrm{subject\ to}\quad \sum_{S\subseteq V}y_S\chi_S\ge w,
    \end{align}
    which is the dual of $\max\Set{w^\top x}{x\in P(f)}$.
    Since $P(f)$ is nonempty and bounded,~\eqref{eq:function-dual} has an optimal solution.
    Among the optimal solutions of~\eqref{eq:function-dual}, choose $y^*$ minimizing $\varphi(y^*)$.

    By \zcref{lem:two-chain}, we can partition $\supp(y^*)$ into two chains $\cC_1,\cC_2$.
    For $i \in \set{1, 2}$, put $p_i\coloneqq\sum_{S\in\cC_i}y_S^*\chi_S$.
    We claim that $\hat f(p_i)=\sum_{S\in\cC_i}f(S)y_S^*$.
    To prove it, fix $i$ and let $p \coloneqq p_i$, let $S_1 \subsetneq \dotsb \subsetneq S_m$ be the nonempty members of $\cC_i$, $S_0 \coloneqq \emptyset$, $t_k\coloneqq\sum_{j=k}^my_{S_j}^*$ for $k \in \set{1, \dotsc, m}$, and $t_{m+1}\coloneqq0$.
    Then, we have $t_1>\cdots>t_m>t_{m+1}=0$, and for $v \in V$,
    \begin{align}
        p_v = \begin{cases}
            t_k & (v \in S_k\setminus S_{k-1},\ k \in \set{1,\ldots,m}),\\
            0 & (v \in V\setminus S_m).
        \end{cases}\notag
    \end{align}
    Hence, in the Lovász extension~\eqref{eq:lovasz} of $f$, we can order $V$ so that each $S_k$ is a prefix.
    Thus, $\hat f(p_i)=\sum_{k=1}^m(t_k-t_{k+1})f(S_k)=\sum_{k=1}^my_{S_k}^*f(S_k)$, as claimed.

    The above argument shows $p_1+p_2 = \sum_{S \subseteq V} y^*_S \chi_S \ge w$ and $\hat f(p_1)+\hat f(p_2)=\sum_{S\subseteq V}f(S)y_S^*=\max_{x\in P(f)}w^\top x$, where the last equality is due to strong duality.
    Conversely, for $x\in P(f)$ and $p\in\R^V_{\ge0}$, we have $p^\top x\le\hat f(p)$ because the coefficients $p_{v_k} - p_{v_{k+1}}$ in~\eqref{eq:lovasz} are nonnegative and $x(S)\le f(S)$ for all $S \subseteq V$.
    Hence, $w^\top x\le w_1^\top x + w_2^\top x\le\hat f(w_1)+\hat f(w_2)$ whenever $w_1,w_2\ge0$ and $w_1+w_2\ge w$.
    This proves~\eqref{eq:function-covering}, with $(p_1,p_2)$ attaining the minimum.

    We now derive~\eqref{eq:function-splitting} from~\eqref{eq:function-covering}.
    For every weight splitting $(w_1,w_2)$ of $w$, the pair $(w_1^+,w_2^+)$ is feasible in~\eqref{eq:function-covering}, as $w_1^+,w_2^+\ge0$ and $w_1^++w_2^+\ge w_1+w_2=w$; hence the minimum in~\eqref{eq:function-splitting} is at least that in~\eqref{eq:function-covering}, which equals $\max_{x\in P(f)}w^\top x$.
    Conversely, let $(p_1,p_2)$ be the minimizer of~\eqref{eq:function-covering} obtained above, and set $w_1\coloneqq p_1$ and $w_2\coloneqq w-p_1$.
    Then, $w_1+w_2=w$, $w_1^+=p_1$ since $p_1\ge0$, and $w_2^+\le p_2$ since $w_2\le p_2$ and $p_2\ge0$.
    Thus, by monotonicity of $\hat f$, we have $\hat f(w_1^+)+\hat f(w_2^+)\le\hat f(p_1)+\hat f(p_2)=\max_{x\in P(f)}w^\top x$, and hence $(w_1,w_2)$ attains the minimum in~\eqref{eq:function-splitting}.

    Next, suppose $B(f)\ne\emptyset$.
    For $x\in B(f)$ and any $p\in\R^V$, we have $p^\top x\le\hat f(p)$, so $w^\top x\le\hat f(w_1)+\hat f(w_2)$ for every weight splitting $(w_1,w_2)$ of $w$.
    Conversely, $B(f)$ is the face of the polytope $P(f)$ maximizing $x(V)$, so for a sufficiently large integer $c$, the maximum of $(w+c\chi_V)^\top x$ over $P(f)$ is attained in $B(f)$ and equals $\max_{x\in B(f)}w^\top x+cf(V)$.
    By~\eqref{eq:function-covering} for $w+c\chi_V$, there are $w_1',w_2'\in\R^V_{\ge0}$ with $w_1'+w_2'\ge w+c\chi_V$ and $\hat f(w_1')+\hat f(w_2')=\max_{x\in B(f)}w^\top x+cf(V)$.
    Then, $w_1\coloneqq w_1'-c\chi_V$ and $w_2\coloneqq w-w_1\le w_2'$ satisfy
    \begin{align}
        \hat f(w_1)+\hat f(w_2)\le\hat f(w_1')-cf(V)+\hat f(w_2')=\max_{x\in B(f)}w^\top x\notag
    \end{align}
    by the translation identity and monotonicity of $\hat f$.
    This proves~\eqref{eq:function-base-splitting}.

    Finally, let $w$ be integer-valued.
    Since the matrix whose columns are the incidence vectors of the members of two chains is totally unimodular~\cite[Theorem~41.11]{schrijver2003combinatorial}, restricting~\eqref{eq:function-dual} to the variables indexed by $\cC_1\cup\cC_2$ yields an integral optimal solution, and the constructions above with this solution and an integer $c$ give integral optimal weight splittings.
\end{proof}

Setting $w$ in~\eqref{eq:function-splitting} to be the all-one vector $\chi_V$ and using the monotonicity of $\hat f$, we obtain
\begin{align}\label{eq:minmax}
    \max_{x\in P(f)}x(V)
    &=\min_{S \subseteq V} \prn[\big]{f(S)+f(V \setminus S)}.
\end{align}

Specializing \zcref{thm:one-third-minmax} to the minimum rank function, we obtain the following.

\begin{corollary}\label{prop:minmax}
    For $i \in \set{1,2}$, let $\bM_i$ be a matroid over $V$ with independent set family $\cI_i$ and base family $\cB_i$.
    Let $\rmin$ be the minimum rank function of $(\bM_1,\bM_2)$.
    For $w\in\R^V$,
    \begin{align}\label{eq:independent-splitting}
        \max_{I \in \cI_1\cap\cI_2} w(I)
        &=\min\Set[\big]{\hatrmin(w_1^+)+\hatrmin(w_2^+)}{w_1,w_2\in\R^V,\ w_1+w_2=w},
    \end{align}
    and, if $\cB_1\cap\cB_2\ne\emptyset$,
    \begin{align}\label{eq:base-splitting}
        \max_{B \in \cB_1\cap\cB_2} w(B)
        =\min\Set[\big]{\hatrmin(w_1)+\hatrmin(w_2)}{w_1,w_2\in\R^V,\ w_1+w_2=w}.
    \end{align}
    If $w$ is integer-valued, both minima are attained by integer-valued vectors.
\end{corollary}

By the greedy algorithm, $\hat r_i(w_i^+) = \max_{I \in \cI_i} w_i(I)$ and $\hat r_i(w_i) = \max_{B \in \cB_i} w_i(B)$ for $i \in \set{1,2}$, where $r_i$ is the rank function of $\bM_i$~\cite{edmonds1971matroids}, implying that the classical weight splitting theorem~\cite{frank1981weighted} has the same form as~\eqref{eq:independent-splitting} and~\eqref{eq:base-splitting}, with the two occurrences of
$\hat r_{\min}$ replaced by $\hat r_1$ and $\hat r_2$, respectively.
Letting $f=\rmin$ in~\eqref{eq:minmax} also gives
\begin{align}
    \max_{I \in \cI_1 \cap \cI_2} |I| &= \min_{S \subseteq V} \prn[\big]{\rmin(S)+\rmin(V \setminus S)}.\notag
\end{align}
This formula, given in~\cite{barasz2025matroid}, is a variant of Edmonds' matroid intersection theorem~\cite{edmonds1970submodular} expressed in terms of $\rmin$, where replacing $\rmin$ with $r_1$ and $r_2$ gives the original one.

\subsection{Computing an Optimal Weight Splitting}\label{sec:dual-computation}

An optimal weight splitting can be computed with the minimum rank oracle, formally stated below.

\begin{theorem}\label{thm:dual}
    Given a maximum-weight common independent set $I^*$, an optimal weight splitting in~\eqref{eq:independent-splitting} can be computed in strongly polynomial time with $\Ord(|I^*|^2n^2)$ additional calls to the minimum rank oracle.
    If both matroids have rank $r$ and a maximum-weight common base $B^*$ is given, the same holds for~\eqref{eq:base-splitting} with $|I^*|=r$.
\end{theorem}

Our approach is outlined as follows.
The proof of \zcref{thm:one-third-minmax} for $f=\rmin$ uses the linear program~\eqref{eq:function-dual} with $f=\rmin$, i.e.,
\begin{align}\label{eq:rank-dual}
    \minimize_{y \in \R_{\ge 0}^{2^V}}\quad \sum_{S\subseteq V}\rmin(S)y_S \quad
    \mathrm{subject\ to}\quad \sum_{S\subseteq V}y_S\chi_S\ge w,
\end{align}
which is the dual of $\max\Set{w^\top x}{x\in P(\rmin)}$.
By \zcref{lem:two-chain}, a solution $y^*$ that minimizes $\varphi(y)$ among all optimal solutions to~\eqref{eq:rank-dual} has support $\supp(y^*)$ equal to the union of two chains $\cC_1,\cC_2$ (cf.~\cite[Remark~4.12]{barasz2025matroid}), and, as shown in the proof of \zcref{thm:one-third-minmax}, $w_1\coloneqq p_1$ and $w_2\coloneqq w-p_1$ form an optimal weight splitting in~\eqref{eq:independent-splitting}, where $p_i\coloneqq\sum_{S\in\cC_i}y^*_S\chi_S$.
As~\eqref{eq:rank-dual} has $2^n$ variables, it remains to compute such a $y^*$.

By complementary slackness, the optimal solutions of~\eqref{eq:rank-dual} are exactly the vectors $y \in \R_{\ge0}^{2^V}$ such that $\supp(y) \subseteq \cF_{I^*}$ (defined in \zcref{sec:tangent-graphs}) and $\sum_{S \subseteq V}y_S\chi_S\in Q_{I^*}\coloneqq\Set{q\in\R^V}{q_v=w_v\ (v\in I^*),\ q_v\ge w_v\ (v\in V\setminus I^*)}$.
We deal with the condition $\supp(y) \subseteq \cF_{I^*}$ via an extended formulation of $\cK_{I^*}$, where $\cK_I$ is the conic hull of $\Set{(\chi_S,1)}{S\in\cF_I}$ for a common independent set $I$.
We show the following.

\begin{lemma}\label{lem:tight-set-cone}
Let $I$ be a common independent set of matroids $\bM_1$ and $\bM_2$.
The cone $\cK_I$ admits a polynomial-size extended formulation with coefficients in $\set{0,\pm1}$.
Such a formulation can be constructed in strongly polynomial time using $\Ord({|I|}^2n^2)$ calls to the minimum rank oracle of $(\bM_1, \bM_2)$.
Moreover, given a feasible point of this formulation, we can express its projection onto $\cK_I$ as a conic combination of polynomially many extreme rays of $\cK_I$ in strongly polynomial time.
\end{lemma}

\begin{proof}
    Let $\bM'_1$ and $\bM'_2$ be the matroids on $V'\coloneqq V+\star$ given by~\eqref{eq:auxiliary} with $k\coloneqq|I|$.
    Then, $B\coloneqq I+\star$ is their common base.
    Construct the hypergraph $H_B$ and bipartite graphs $G_0$ and $G_e$ $(e\in F \coloneqq E(G_0) \setminus E(H_B))$ defined in \zcref{sec:exch-hypergraphs} for $(\bM_1', \bM_2')$ and $B$ with $\Ord(k^2n^2)$ oracle calls.
    For $S\subseteq V$, as $|B\cap S|=|I\cap S|\le k$, \eqref{eq:auxiliary-oracle} shows that $S\in\cF_I$ if and only if $\rminp(S)=|B\cap S|$, that is, $S\symdif B\in\cS(H_B)$ by \zcref{lem:rank,lem:hypergraph}.
    Since $\star\in B$, a set $Z\subseteq V'$ satisfies $B\symdif Z\subseteq V$ if and only if $\star\in Z$; hence, \zcref{lem:decomposition} gives
    \begin{align}
        \cF_I &= \Set{B\symdif S}{S\in\cS(G_0),\ \star\in S} \cup\bigcup_{e\in F}\Set{B\symdif(S\cup e)}{S\in\cS(G_e),\ \star\in S\cup e}.\label{eq:dual-tight-lift}
    \end{align}
    To treat $G_0$ and $G_e$ uniformly, let $J\coloneqq\set{0}\cup F$, $C_0\coloneqq\emptyset$, and $C_e\coloneqq e$ $(e\in F)$.
    Then,~\eqref{eq:dual-tight-lift} can be expressed as
    \begin{align}\label{eq:dual-tight-lift2}
        \cF_I = \bigcup_{j\in J}\Set{B\symdif(S \cup C_j)}{S\in\cS(G_j),\ \star\in S \cup C_j}.
    \end{align}

    For $j\in J$, define
    \begin{align}
        P_j\coloneqq\Set*{x\in[0,1]^{V'}}{\begin{array}{ll}
            x_u+x_v\le1 & (\set{u,v}\in E(G_j)),\\
            x_v=1 & (v\in C_j+\star),\\
            x_v=0 & (v\in V'\setminus(V(G_j) \cup C_j))
        \end{array}}.\notag
    \end{align}
    The integral points of $P_j$ are the vectors $\chi_{S \cup C_j}$ with $S\in\cS(G_j)$ and $\star\in S \cup C_j$, and $P_j$ is their convex hull since $G_j$ is bipartite~\cite[Theorem~19.7]{schrijver2003combinatorial}.
    Define the polyhedral cone
    \begin{align}
        \hat P_j\coloneqq\Set{(z,\lambda)\in\R^{V'}\times\R_{\ge0}}{z\in\lambda P_j},\notag
    \end{align}
    which is the conic hull of the vectors $(\chi_{S \cup C_j},1)$ with $S\in\cS(G_j)$ and $\star\in S \cup C_j$.
    For any $S\subseteq V'$, we have $\chi_{B\symdif S}=\chi_B+\sigma_B(\chi_S)$, where $\sigma_B\colon\R^{V'}\to\R^{V'}$ is defined in \zcref{sec:tangent-graphs}.
    Thus, the linear map $L\colon\R^{V'}\times\R\to\R^V\times\R$ defined by $L(z,\lambda)\coloneqq((\lambda\chi_B+\sigma_B(z))|_V,\ \lambda)$ sends $(\chi_S,1)$ to $(\chi_{B\symdif S},1)$ for every $S\subseteq V'$ with $\star\in S$.
    Hence,~\eqref{eq:dual-tight-lift2} implies
    \begin{align}
        \cK_I=\Set*{\sum_{j\in J}L(z^j,\lambda^j)}{(z^j,\lambda^j)\in\hat P_j\ (j\in J)}.\notag
    \end{align}
    Explicitly, $\cK_I$ is the projection onto the $(q,m)$-coordinates of the set of vectors $(q,m,(z^j,\lambda^j)_{j\in J})$ with $q\in\R^V$, $m\in\R$, $z^j\in\R^{V'}$, and $\lambda^j\in\R$ satisfying
    \begin{alignat}{2}
        &q_v=\textstyle\sum_j(\lambda^j-z^j_v)\quad(v\in I),&\quad&q_v=\textstyle\sum_jz^j_v\quad(v\in V\setminus I),\qquad m=\textstyle\sum_j\lambda^j,\label{eq:tight-cone-ef1}\\
        &0\le z^j_v\le\lambda^j\quad(j\in J,\ v\in V'),&\quad&z^j_u+z^j_v\le\lambda^j\quad(j\in J,\ \set{u,v}\in E(G_j)),\label{eq:tight-cone-ef2}\\
        &z^j_v=\lambda^j\quad(j\in J,\ v\in C_j+\star),&\quad&z^j_v=0\quad(j\in J,\ v\notin V(G_j)\cup C_j),\label{eq:tight-cone-ef3}
    \end{alignat}
    where~\eqref{eq:tight-cone-ef1} expresses $(q,m)=\sum_{j\in J}L(z^j,\lambda^j)$, and~\eqref{eq:tight-cone-ef2} and~\eqref{eq:tight-cone-ef3} express $(z^j,\lambda^j)\in\hat P_j$.
    This system has $\Ord(|J||V'|)=\Ord(kn^2)$ variables and $\Ord(|J||B||V'|)=\Ord(k^2 n^2)$ constraints, and all coefficients are in $\set{0,\pm1}$.

    Finally, we explain the conic decomposition of a feasible solution $(q, m, (z^j,\lambda^j)_{j\in J})$ of this system.
    For each $j \in J$ with $\lambda^j > 0$, the vector $z^j/\lambda^j \in P_j$ can be decomposed in strongly polynomial time into a convex combination of at most $|V'|+1 = \Ord(n)$ vertices of $P_j$~\cite[Theorem~5.15(ii)]{schrijver2003combinatorial}.
    Multiplying these decompositions by $\lambda^j$ and applying $L$ yields the desired conic combination of $(q,m) \in \cK_I$ with $\Ord(kn^2)$ terms, each of which spans an extreme ray of $\cK_I$.
\end{proof}

We now prove \zcref{thm:dual} using \zcref{lem:tight-set-cone}.

\begin{proof}[Proof of \zcref{thm:dual}]
    By \zcref{lem:two-chain} and the proof of \zcref{thm:one-third-minmax}, it suffices to compute an optimal solution of~\eqref{eq:rank-dual} that minimizes $\varphi(y)$ among all optimal solutions.
    By complementary slackness, this problem is equivalent to
    \begin{align}\label{eq:lex-dual}
        \minimize_{y \in \R_{\ge 0}^{\cF_{I^*}}}\quad \sum_{S\in \cF_{I^*}} (n^2 - {|S|}^2) y_S \quad
        \text{subject\ to} \quad \sum_{S\in \cF_{I^*}} y_S \chi_S \in Q_{I^*}.
    \end{align}
    To linearize the coefficient $n^2-{|S|}^2$ of $y_S$ in~\eqref{eq:lex-dual} with respect to $|S|$, we rewrite~\eqref{eq:lex-dual} as
    \begin{align}\label{eq:copy-lp}
        \minimize_{y^0,\dotsc,y^n\in\R_{\ge0}^{\cF_{I^*}}}\quad \sum_{k=0}^n\sum_{S\in\cF_{I^*}}(n^2+k^2-2k|S|)y^k_S \quad
        \text{subject\ to} \quad \sum_{k=0}^n\sum_{S\in\cF_{I^*}}y^k_S\chi_S\in Q_{I^*}.
    \end{align}
    These problems have the same optimal value.
    Indeed, $n^2+k^2-2k|S| \ge n^2 - {|S|}^2$ holds, with equality if $k=|S|$.
    Hence, setting $y^k_S\coloneqq y_S$ if $|S|=k$ and $y^k_S\coloneqq0$ otherwise turns a feasible solution $y$ of~\eqref{eq:lex-dual} into a feasible solution of~\eqref{eq:copy-lp} with the same objective value.
    Conversely, for a feasible solution ${(y^k)}_k$ of~\eqref{eq:copy-lp}, the vector $y\coloneqq\sum_{k=0}^n y^k$ is feasible for~\eqref{eq:lex-dual}, and $\varphi(y)$ is at most the objective value of ${(y^k)}_k$ since $y^k\ge0$.
    In particular, $\sum_{k=0}^n y^k$ is optimal for~\eqref{eq:lex-dual} whenever ${(y^k)}_k$ is optimal for~\eqref{eq:copy-lp}.

    For each $k$, put $q^k \coloneqq \sum_{S \in \cF_{I^*}} y^k_S \chi_S$ and $m^k \coloneqq \sum_{S \in \cF_{I^*}} y^k_S$.
    Then,
    \begin{align}
        \sum_{S\in\cF_{I^*}}(n^2+k^2-2k|S|)\,y^k_S=(n^2+k^2)m^k-2kq^k(V),\notag
    \end{align}
    and the constraint of~\eqref{eq:copy-lp} reads $\sum_{k=0}^n q^k\in Q_{I^*}$.
    Thus, using $\cK_{I^*}$, we can rewrite~\eqref{eq:copy-lp} as
    \begin{align}
        \begin{aligned}\label{eq:uncrossing-lp}
            &\minimize\quad&& \sum_{k=0}^n\bigl((n^2+k^2)m^k-2kq^k(V)\bigr)\\
            &\text{subject\ to}\quad&& (q^k,m^k)\in\cK_{I^*}\ (k=0,\dotsc,n),\quad \sum_{k=0}^n q^k\in Q_{I^*}.
        \end{aligned}
    \end{align}

    We construct the extended formulation of $\cK_{I^*}$ in \zcref{lem:tight-set-cone} with $\Ord(|I^*|^2n^2)$ oracle calls and use a copy of it for each $(q^k,m^k)$.
    This turns~\eqref{eq:uncrossing-lp} into a linear program of polynomial size with a $\set{0,\pm1}$ constraint matrix.
    Since~\eqref{eq:lex-dual} has an optimal solution, so does~\eqref{eq:uncrossing-lp} by the equivalences above.
    Hence, by \zcref{prop:lp}, we can compute an optimal solution of~\eqref{eq:uncrossing-lp} in strongly polynomial time.
    By \zcref{lem:tight-set-cone}, we decompose each $(q^k,m^k) \in \cK_{I^*}$ of this optimal solution as a conic combination $\sum_{S\in\cF_{I^*}}y^k_S(\chi_S,1)$ with polynomially many nonzero $y^k_S$.
    Then, ${(y^k)}_k$ is optimal for~\eqref{eq:copy-lp}, and hence $y\coloneqq\sum_{k=0}^n y^k$ is optimal for~\eqref{eq:lex-dual}; that is, $y$ minimizes $\varphi$ among all optimal solutions of~\eqref{eq:rank-dual}.

    For the second statement, let $w'\coloneqq w+c\chi_V$ with $c\coloneqq1+\sum_{v\in V}|w_v|$, so that $B^*$ is a maximum-weight common independent set with respect to $w'$.
    Applying the above to $B^*$ and $w'$ gives $p_1,p_2$ with $p_1+p_2\ge w'$ and $\hat r_{\min}(p_1)+\hat r_{\min}(p_2)=w(B^*)+cr$.
    For $w_1\coloneqq p_1$ and $w_2\coloneqq w-p_1$, we have $w_2+c\chi_V\le p_2$, so the monotonicity of $\hat r_{\min}$ and the translation property $\hat r_{\min}(p+c\chi_V)=\hat r_{\min}(p)+cr$ give $\hat r_{\min}(w_1)+\hat r_{\min}(w_2)\le w(B^*)$, and equality holds by~\eqref{eq:base-splitting}.
\end{proof}

\begin{remark}
    The system~\eqref{eq:tight-cone-ef1},~\eqref{eq:tight-cone-ef2}, and~\eqref{eq:tight-cone-ef3} is closely related to the LP dual of the system in \zcref{thm:ef} for $(\bM'_1,\bM'_2)$ and $B$.
    Consider the linear program maximizing $c^\top d$ for $c\in\R^{V'}$ subject to the system in \zcref{thm:ef}.
    Its dual has variables $\alpha\in\R$ for $d(V')=0$, $\pi^0\in\R^{V'}_{\ge0}$ and $\lambda^0\ge0$ for~\eqref{eq:ef-k-capacity} and~\eqref{eq:ef-k}, and $\pi^e\in\R^{V(G_e)}_{\ge0}$ and $\lambda^e\ge0$ for~\eqref{eq:ef-q-capacity} and~\eqref{eq:ef-q} for $e\in F$.
    Since the primal problem is homogeneous, the dual is the feasibility problem
    \begin{align}
        &\alpha\chi_{V'}+\sum_{j\in J}\bigl(\lambda^j\chi_{V'\setminus B}-\sigma_B(\pi^j+\lambda^j\chi_{X_j})\bigr) = c,\label{eq:ef-dual1}\\
        &\pi^j_u+\pi^j_v\ge\lambda^j\quad(j\in J,\ \set{u,v}\in E(G_j)),\label{eq:ef-dual2}
    \end{align}
    where $X_e$ $(e\in F)$ is defined in \zcref{sec:exch-hypergraphs}, $X_0\coloneqq\emptyset$, and each $\pi^j$ is extended to $V'$ by zero.
    By LP duality, the polar cone $T_B^\circ\coloneqq\Set{c\in\R^{V'}}{c^\top d\le0\ (d\in T_B)}$ of the tangent cone $T_B$ consists of the vectors $c$ for which the system~\eqref{eq:ef-dual1} and~\eqref{eq:ef-dual2} has a solution $(\alpha,(\pi^j,\lambda^j)_{j\in J})$.
    Letting $z^j\coloneqq\lambda^j\chi_{V'\setminus X_j}-\pi^j$ and using $\sigma_B(\chi_{V'})=\chi_{V'\setminus B}-\chi_B$, the equation~\eqref{eq:ef-dual1} becomes $\alpha\chi_{V'}+\sum_{j\in J}(\lambda^j\chi_B+\sigma_B(z^j))=c$, whose restriction to $V$ with $\alpha=0$ is the equations for $q=c|_V$ in~\eqref{eq:tight-cone-ef1}.
    Moreover,~\eqref{eq:ef-dual2} and the non-negativity constraints for $\pi^j$ and $\lambda^j$ correspond to~\eqref{eq:tight-cone-ef2} and~\eqref{eq:tight-cone-ef3}, except that $z^j\ge0$ is dropped and $C_j+\star$ is replaced by $C_j$.
    In this sense, \zcref{lem:tight-set-cone} is a polar counterpart of \zcref{thm:ef}.
\end{remark}

\section*{Acknowledgments}

We used GPT-6 Astra to assist with the proofs of \zcref{thm:main,thm:dual}.
All arguments were independently verified by the author, who takes full responsibility for its content.

\printbibliography

\end{document}